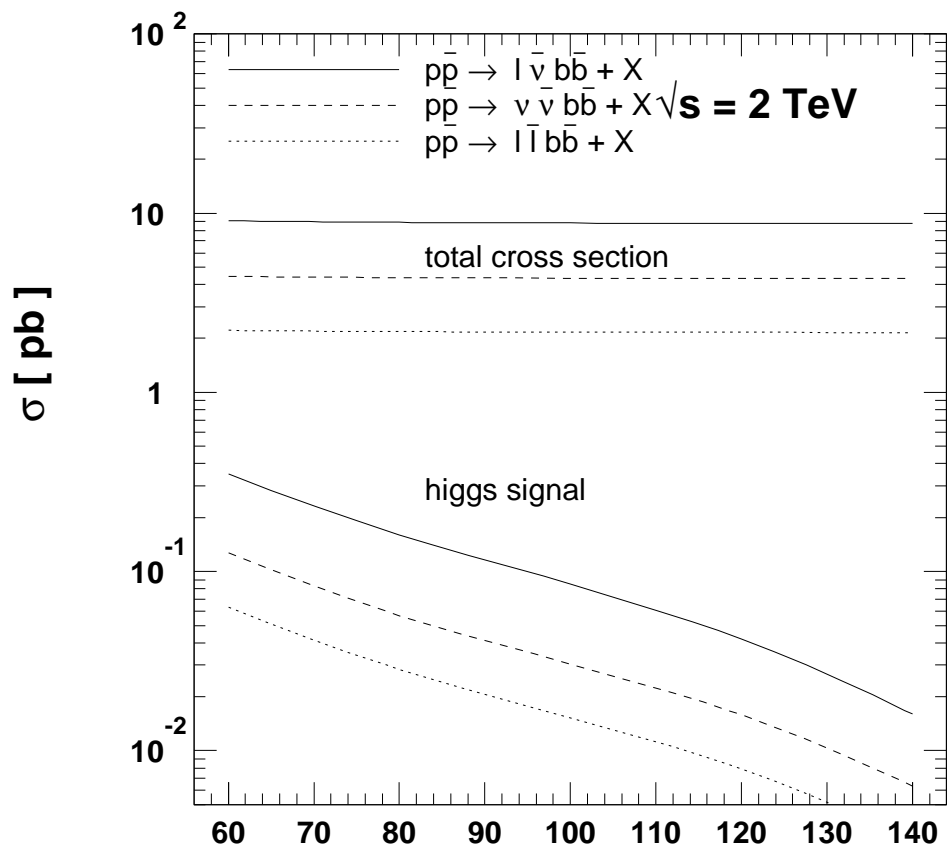
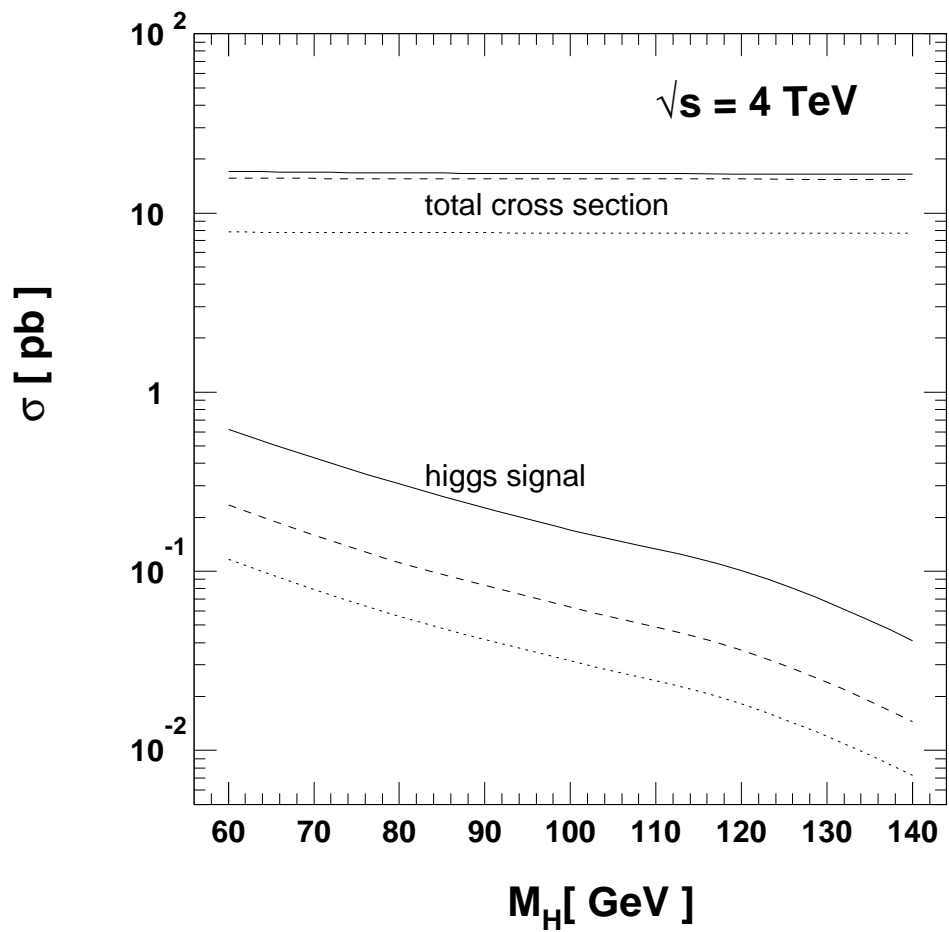

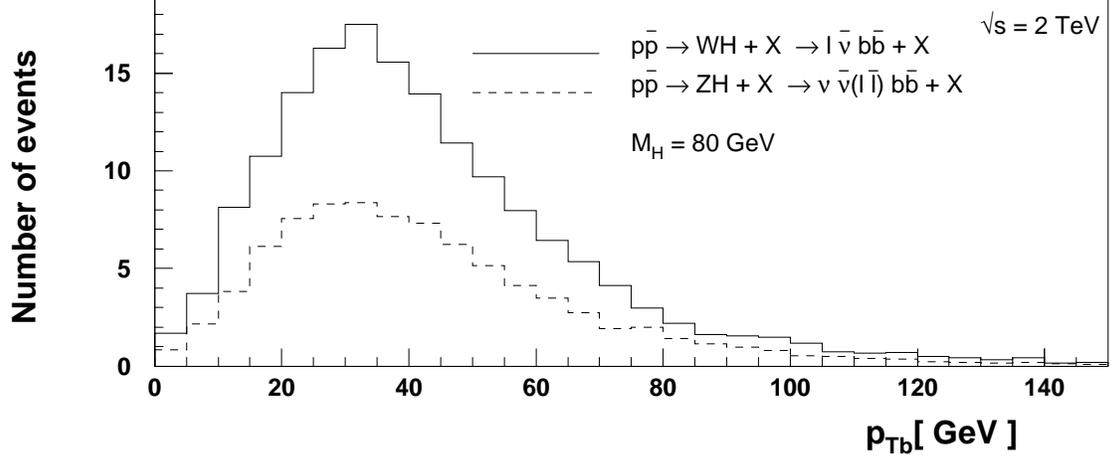
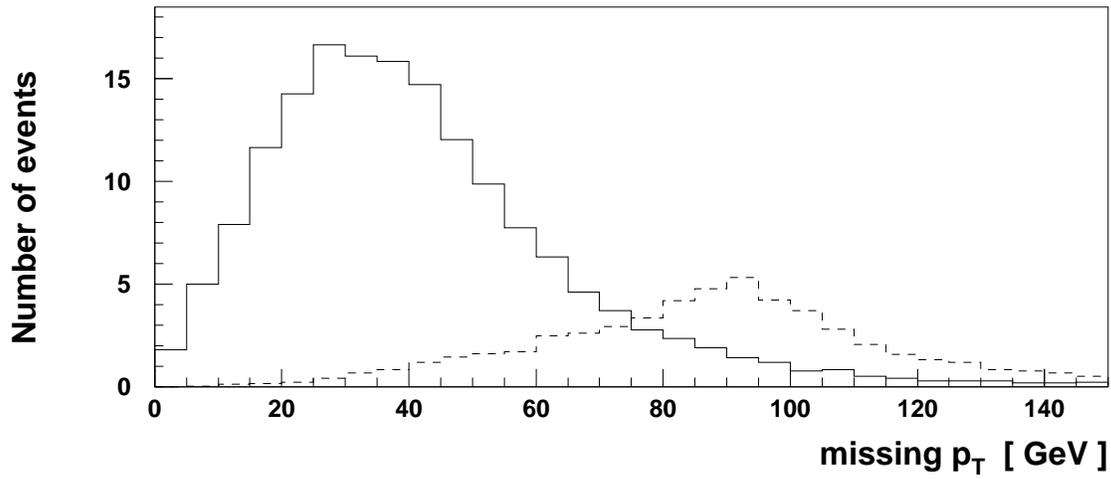
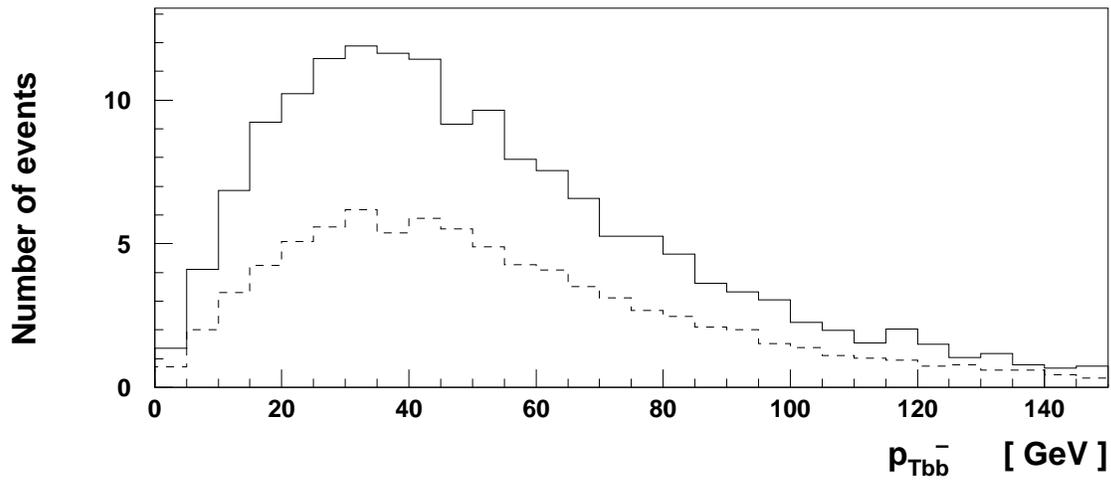
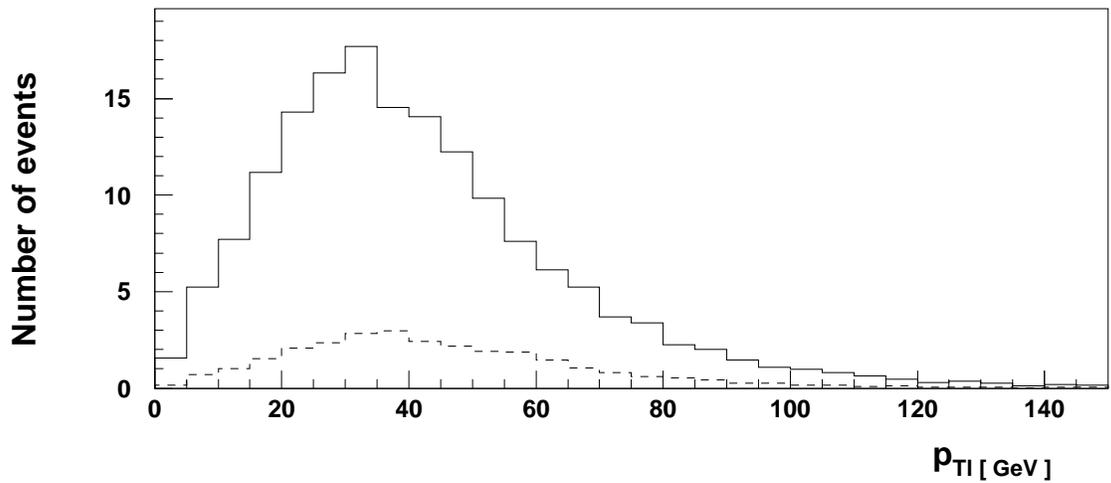

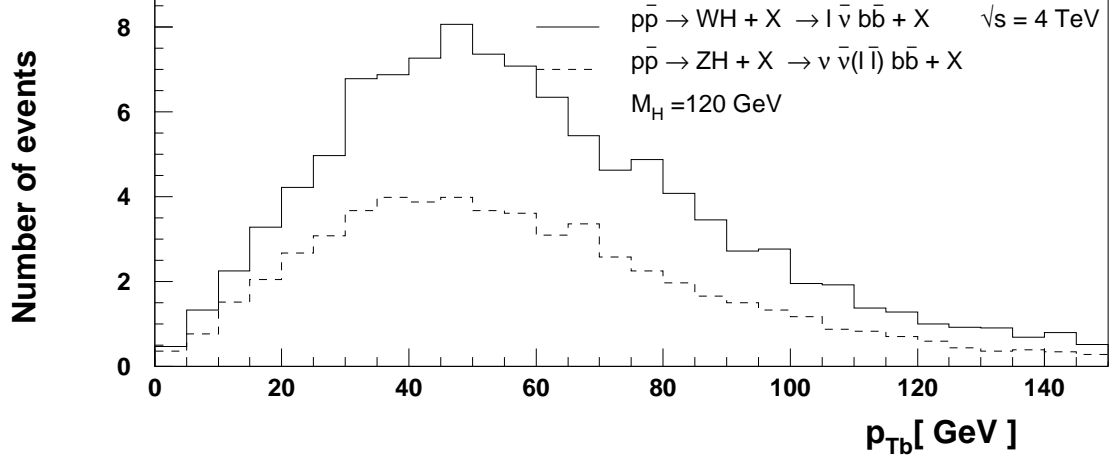

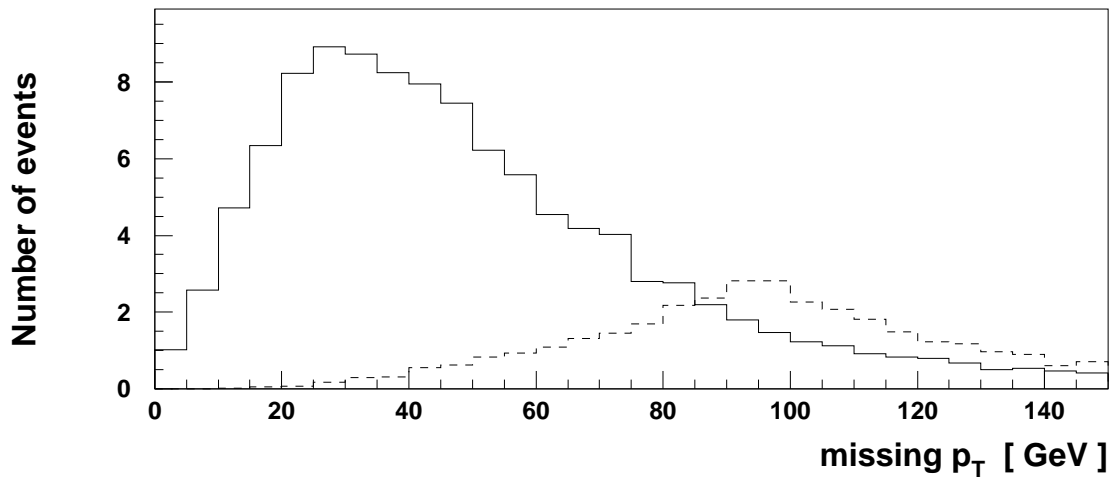

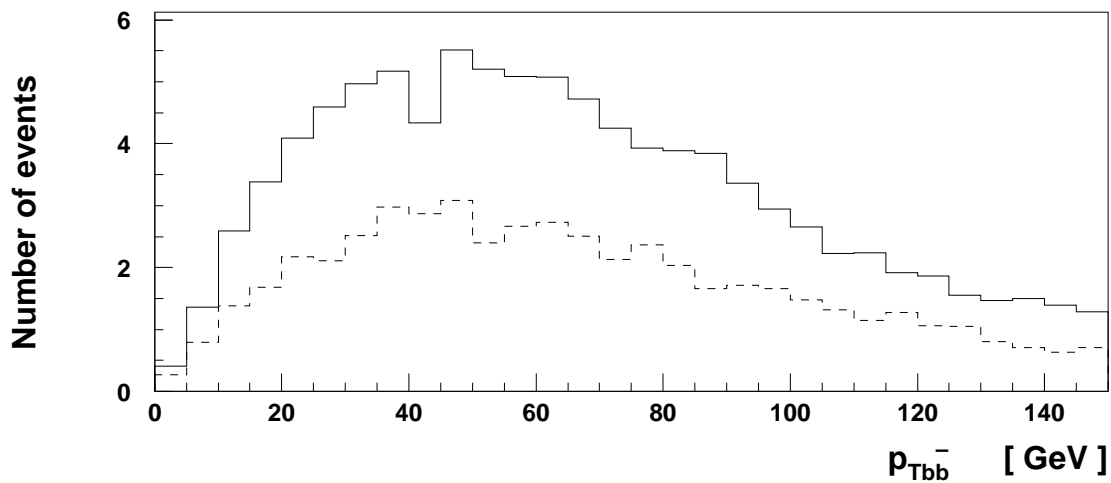

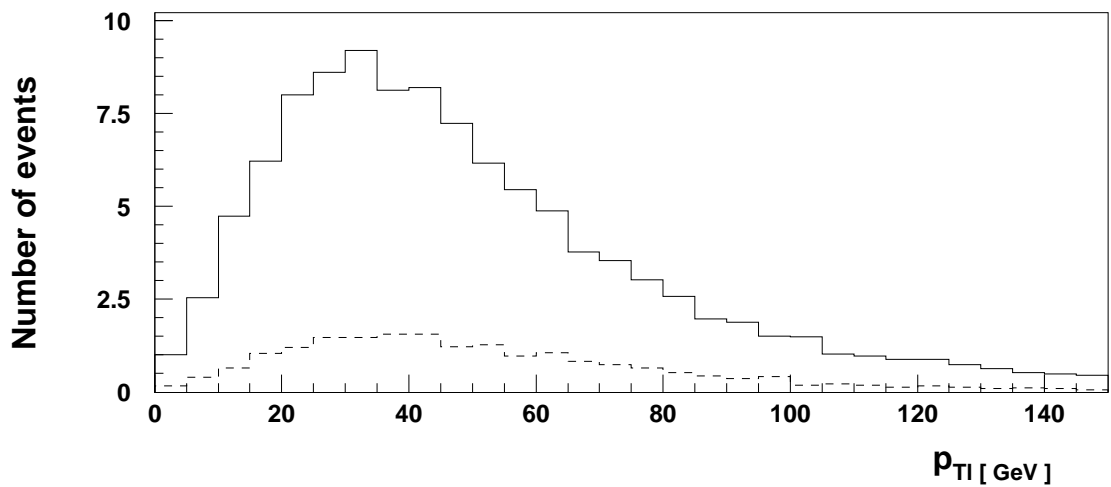

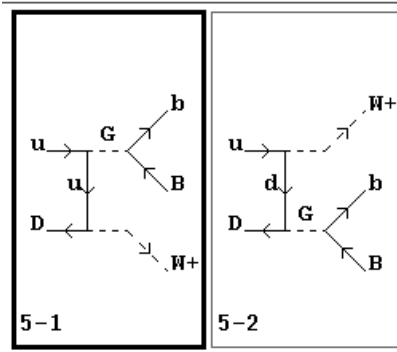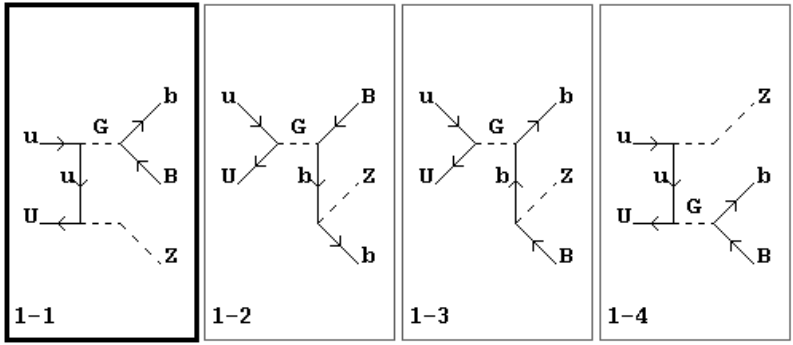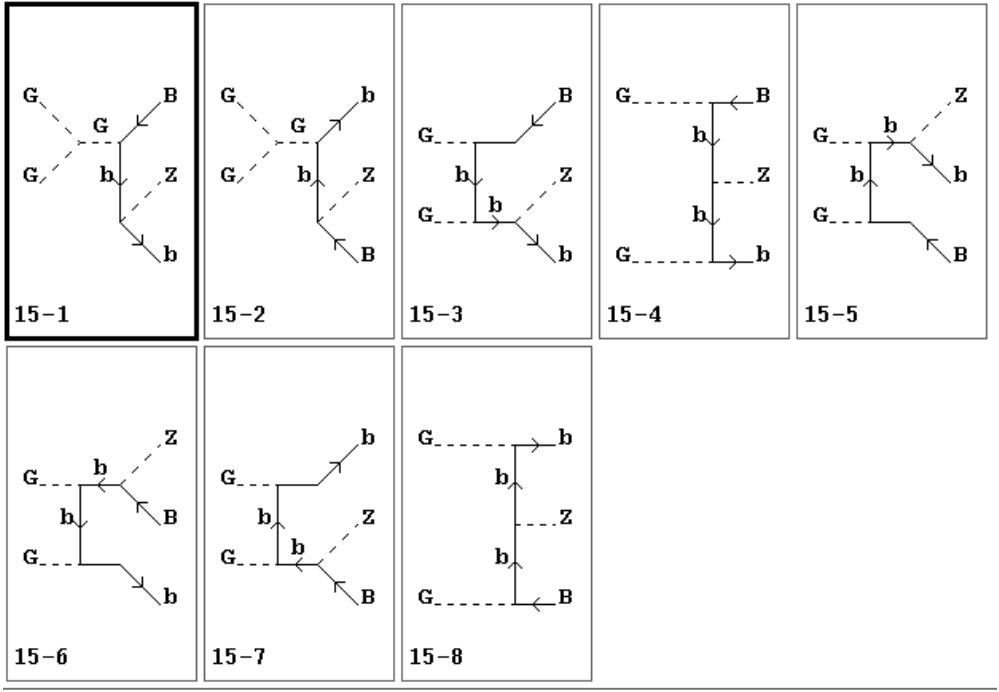

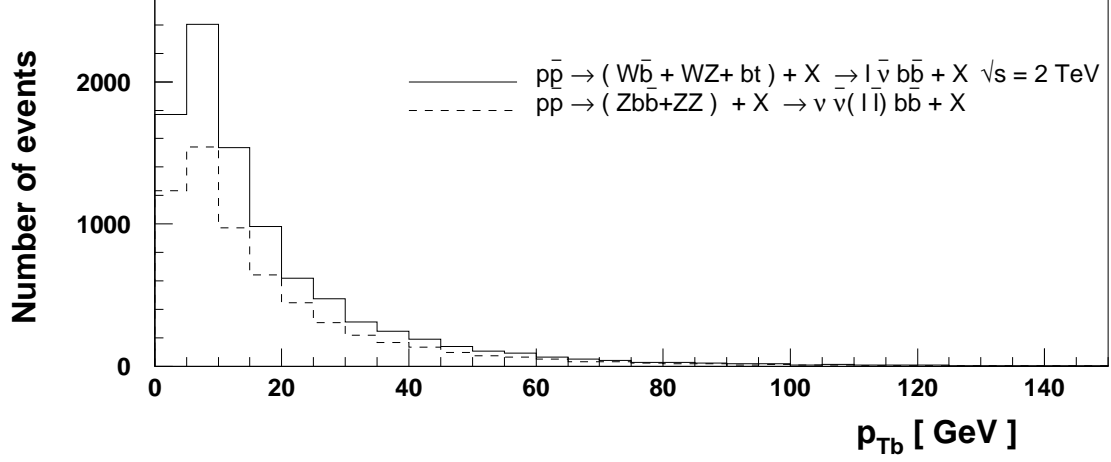
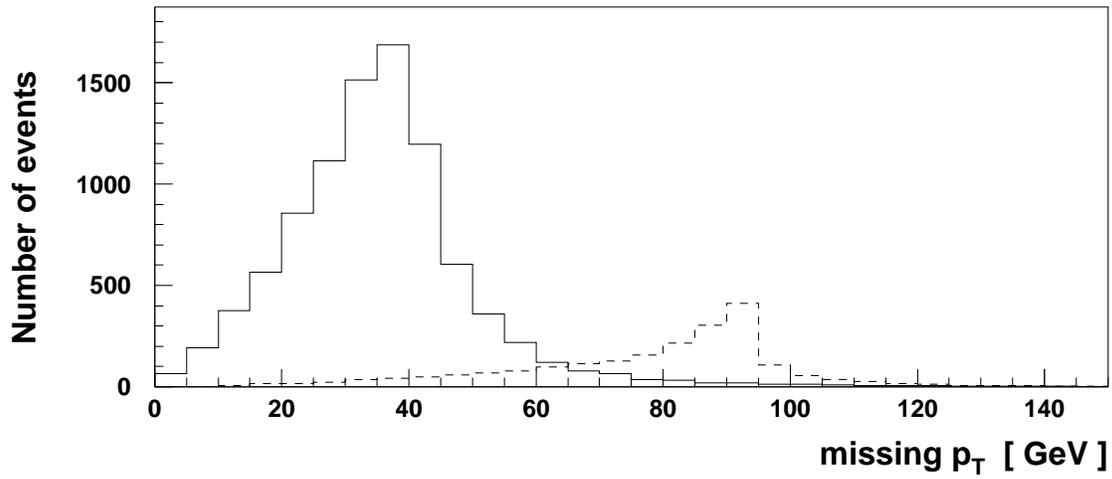
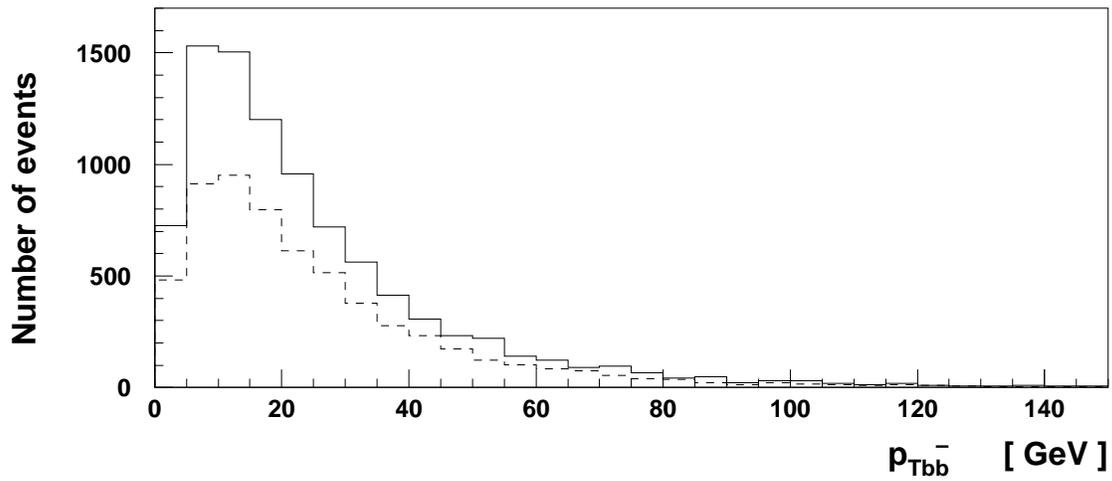
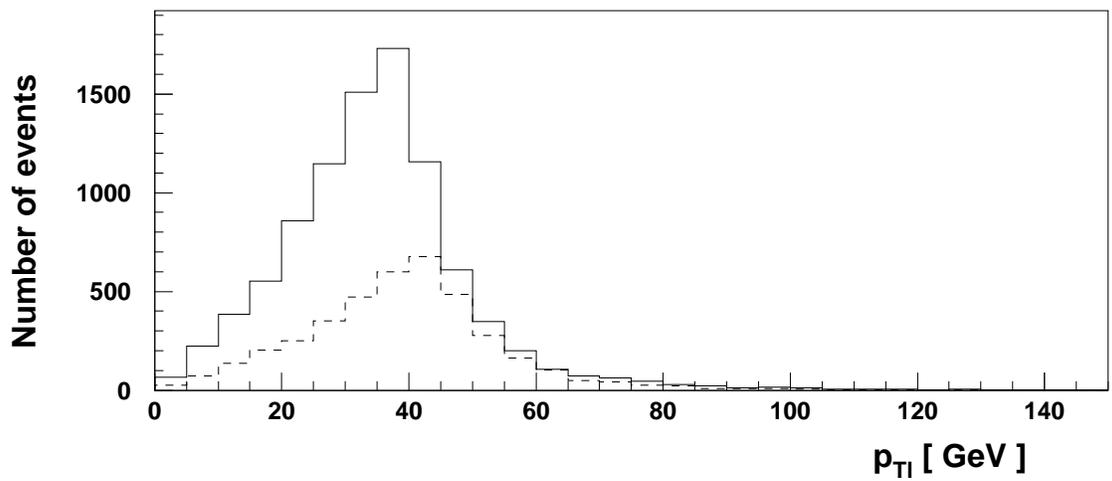

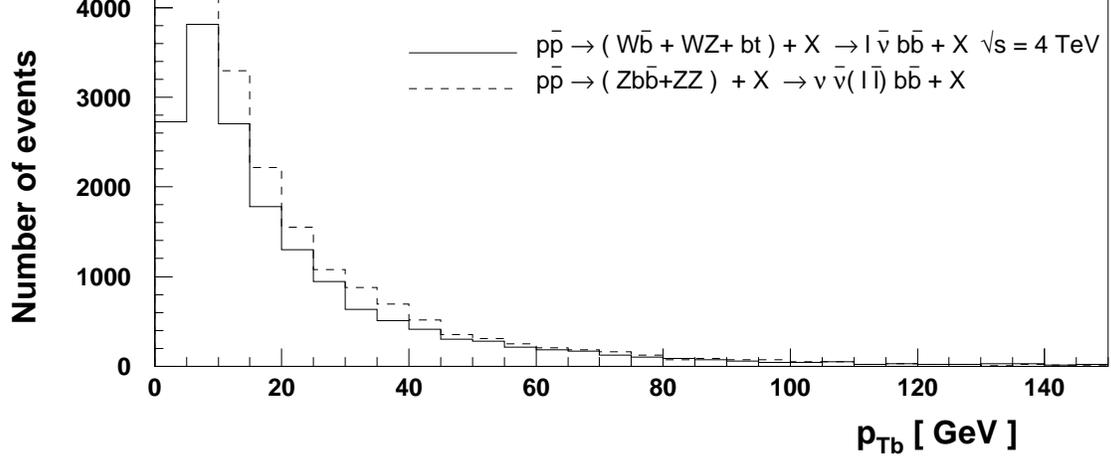
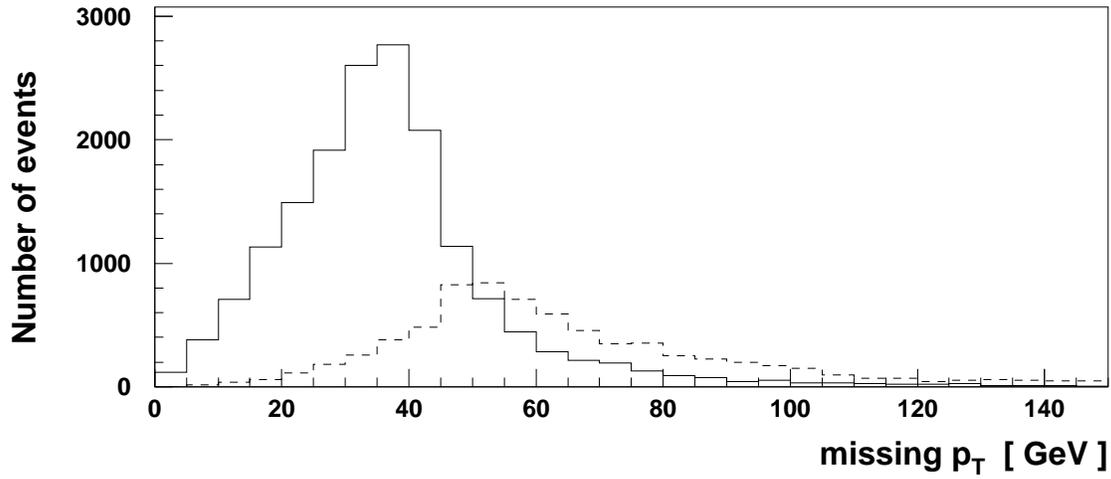
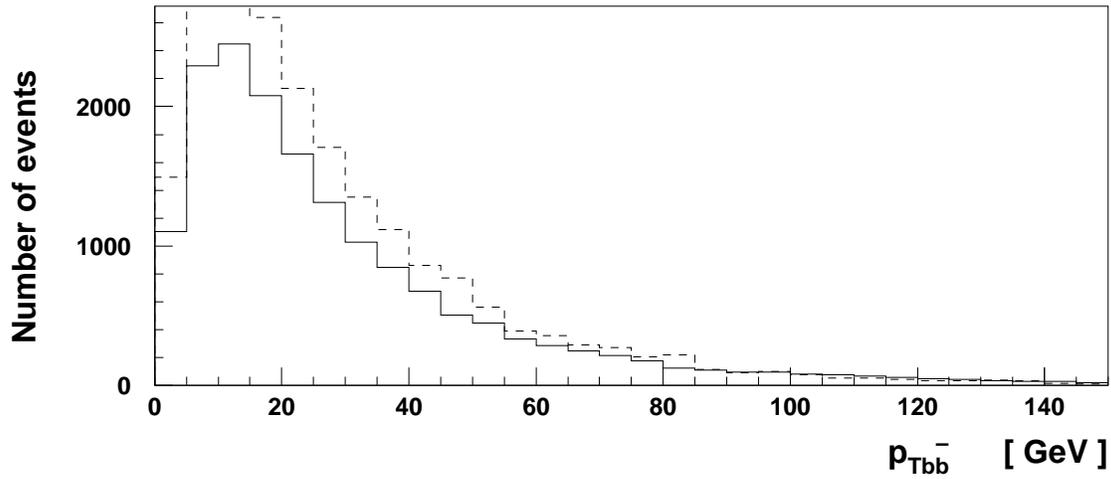
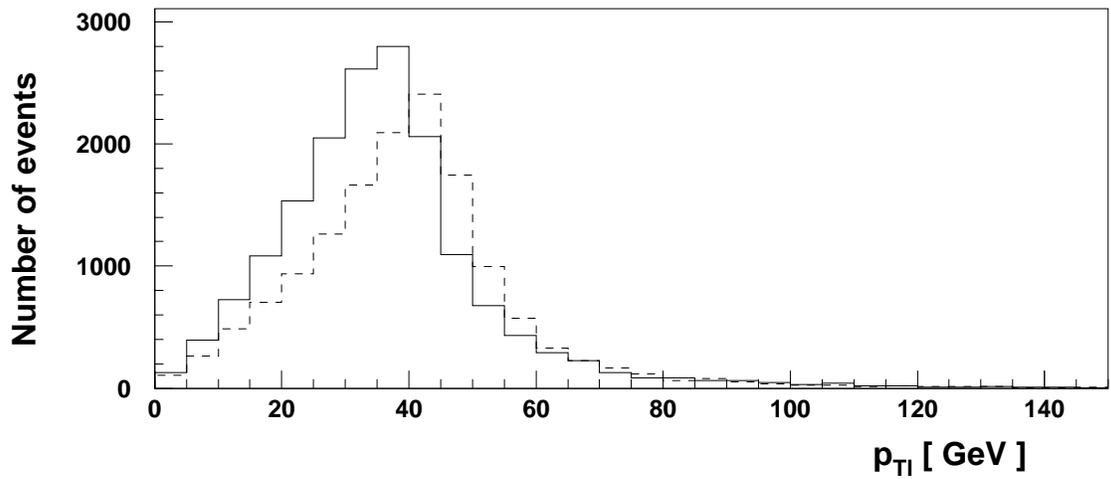

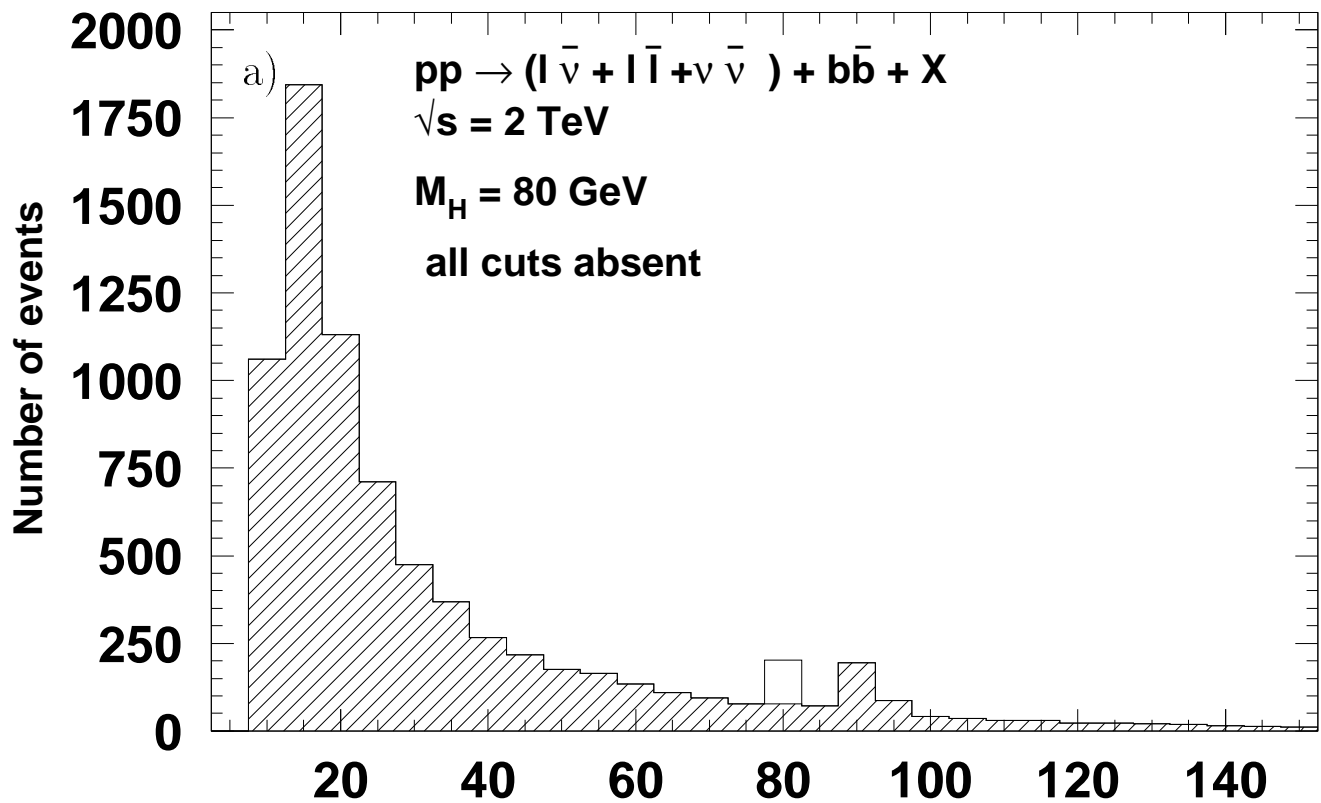
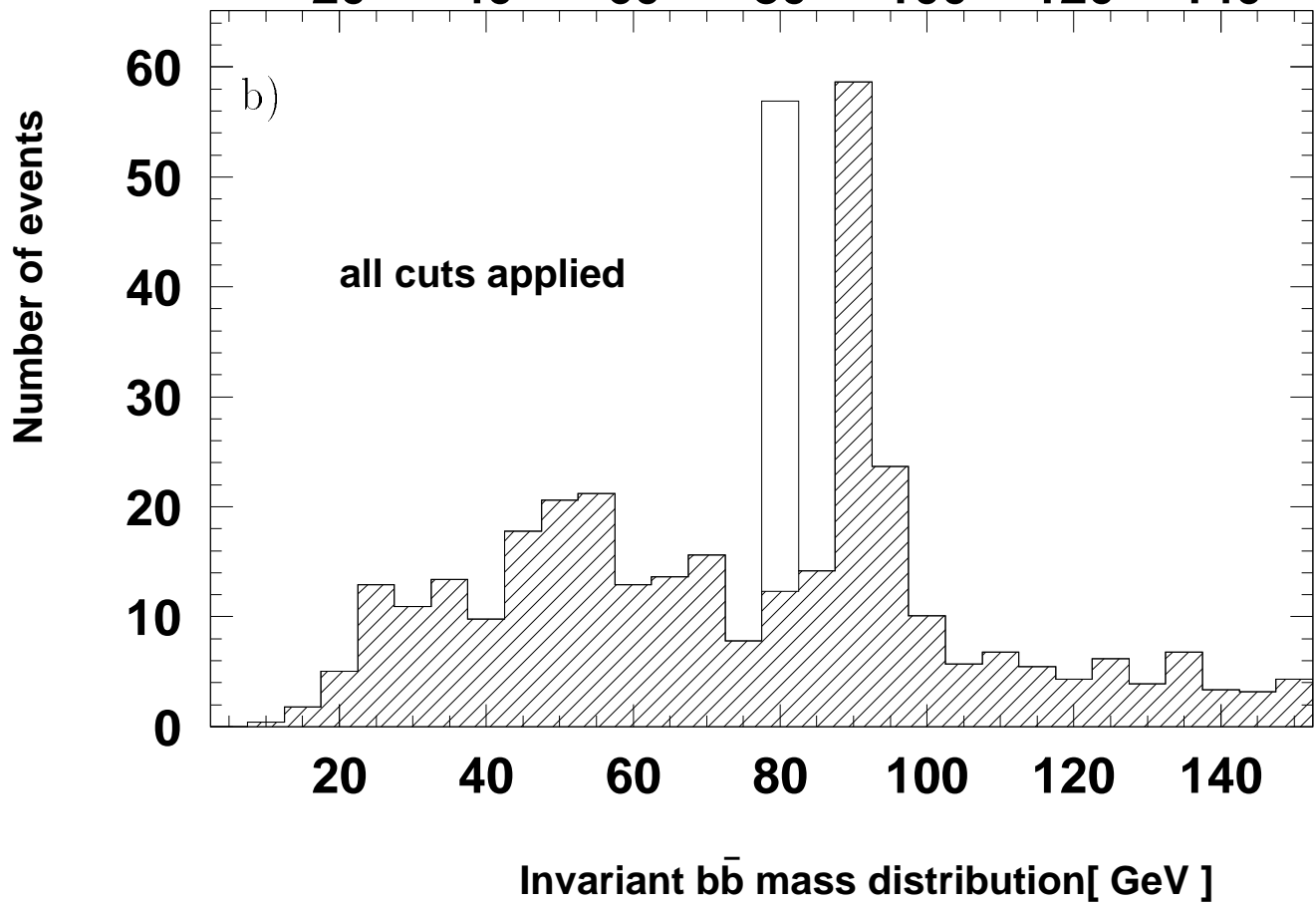

Invariant b$\bar{\text{b}}$ mass distribution [ GeV ]

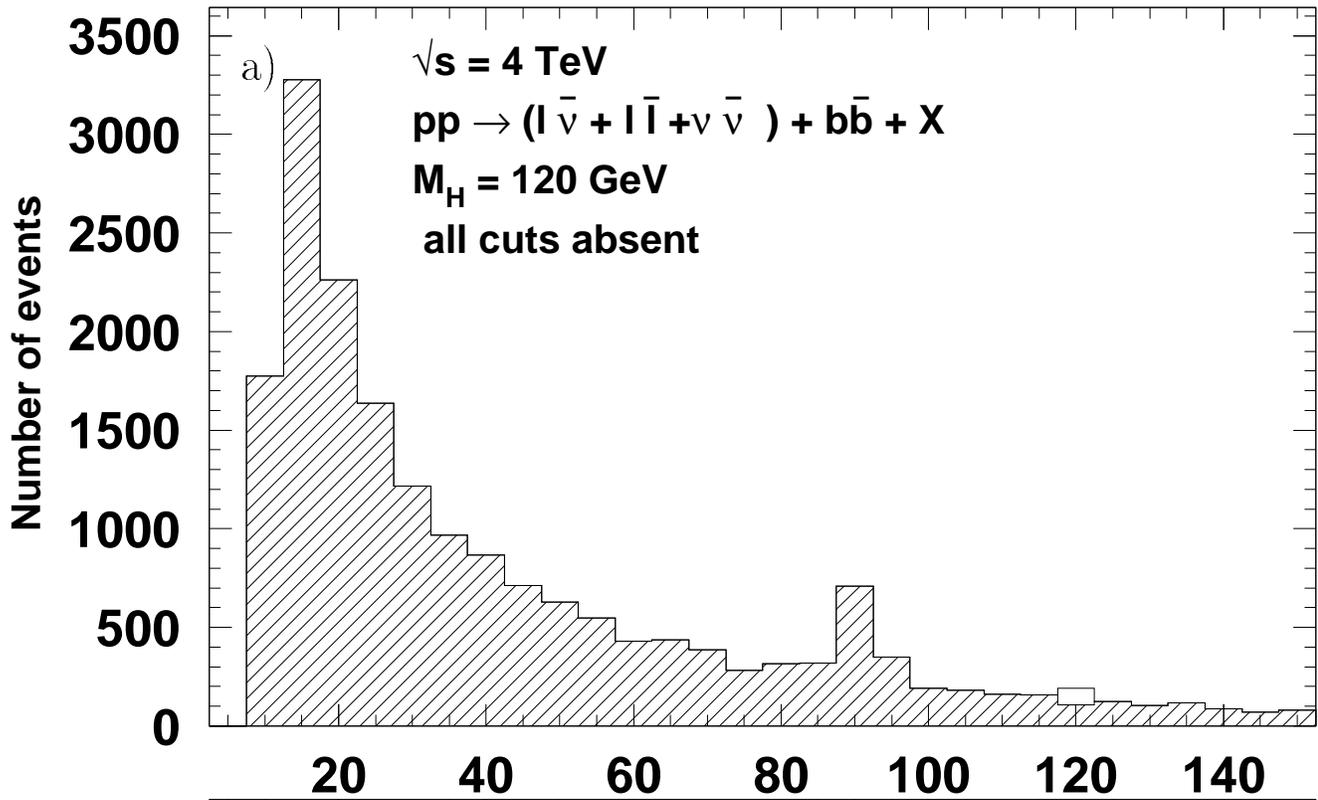
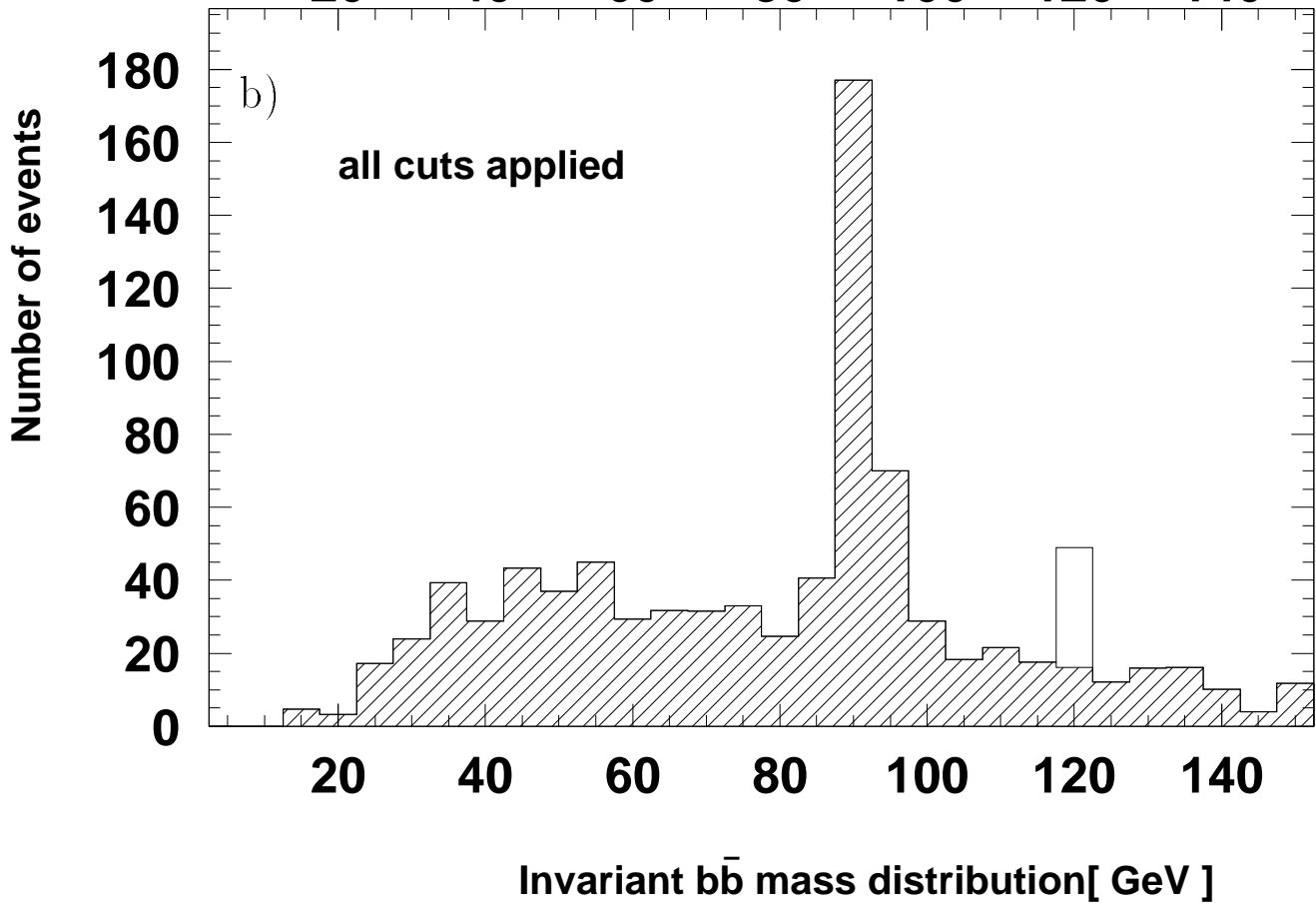

Invariant b̄b mass distribution[ GeV ]

# LIGHT AND INTERMEDIATE HIGGS BOSON SEARCH AT TEVATRON ENERGIES

A. Belyaev, E. Boos, L. Dudko

*Nuclear Physics Institute, Moscow State University,*

*119899, Moscow, Russia*

## 1  Introduction

The forthcoming luminosity upgrade of the Fermilab TEVATRON $p\bar{p}$ collider up to least 500 pb$^{-1}$ per year and the installation of the precise tracking systems in the existing detectors allowing the detection of the secondary vertices from b-quarks ( b-tagging) will open new opportunities for the Higgs boson search at the TEVATRON.

In [1, 2] it was shown that at the TEVATRON energies ($\sqrt{s} = 2$ or 4 TeV) the most promising will be Higgs boson search in the processes of Higgs production in association with the electroweak W and Z bosons. These processes are the direct analogue of the bremsstrahlung mechanism of the Higgs production in $e^+e^-$ collisions. Calculations of the Higgs signal reactions

$$p\bar{p} \to W^{\pm}H + X, \tag{1}$$

$$p\bar{p} \to ZH + X \tag{2}$$

have been done for LHC and SSC energy range in a number of papers [3]. In the paper [1] the calculation of the reaction (1) as well as complete tree level calculation of the reaction $p\bar{p} \to Wb\bar{b} + X$ have been done at TEVATRON energies. In the paper [2] the reactions (1) and (2) at TEVATRON energies have been studied including QCD radiative corrections.

In this paper we present the more detailed study of the Higgs signal processes in the Higgs mass range $M_H \leq 140$ GeV where the Higgs decay into a pair of $b\bar{b}$ quarks dominates, as well as the study of the basic background processes including an analysis of the 4-fermion final state distributions.

The paper is organized as follows. In part 2 we will present the results for the total rate of the 4-fermion final state reactions. The rates, most promising signatures and various kinematical distributions of the Higgs production processes (1, 2) are analyzed. In part 3 we present the results of the background simulation. The optimal kinematical cuts are proposed based on the comparison of the signal and background distributions. The effective mass distribution of the bb-pair after applying cuts and the extraction of the Higgs signal from the background are discussed.

## 2  Total event rate and the Higgs boson contribution

Due to the fact that the dominating decay mode of the Higgs in the mass interval $M_H \leq 140$ GeV is $H \to b\bar{b}$ the Higgs production reactions (1), (2) lead to the following parton final states:

$$p\bar{p} \to b\bar{b}l^{\pm}\nu + X \tag{3}$$
$$p\bar{p} \to b\bar{b}\nu\bar{\nu} + X \tag{4}$$
$$p\bar{p} \to b\bar{b}l^+l^- + X \tag{5}$$
$$p\bar{p} \to b\bar{b}q\bar{q}(\bar{q}') + X, \tag{6}$$

where $l$ is one of the charged lepton $e, \mu, \tau$.

In the paper we will consider in detail the reactions (3)-(5) which correspond to the leptonic decay modes of W- and Z-bosons. These reactions provide more clean event signatures for detection, namely

$$p\bar{p} \rightarrow 2b\text{-}jets + l + \not{p}_\perp(\not{E}_\perp) + X \qquad (7)$$
$$p\bar{p} \rightarrow 2b\text{-}jets + \not{p}_\perp(\not{E}_\perp) + X \qquad (8)$$
$$p\bar{p} \rightarrow 2b\text{-}jets + l^+l^- + X, \qquad (9)$$

where $\not{p}_\perp(\not{E}_\perp)$ denotes missing transverse momentum (energy).

The reaction (6) leads to the signature

$$p\bar{p} \rightarrow 2b\text{-}jets + 2\text{-}jets + X \qquad (10)$$

with the largest event rate. However the reaction (6) is more complicated phenomenologically because the complete $2 \rightarrow 4$ calculation has to be done in this case, and experimentally because the corresponding signature (10) is more difficult for separation of the Higgs signal contribution from the background one. A detailed study of the reaction (9) will be done in a separate paper.

The complete set of tree level Standard Model diagrams contributing to the final states (3), (4), (5) includes the following numbers of $2 \rightarrow 4$ diagrams:

18 diagrams for the parton reaction $q\bar{q} \rightarrow b\bar{b}l^\pm\nu$;
33 diagrams for the reaction $q\bar{q} \rightarrow b\bar{b}l^+l^-$;
16 diagrams for the reaction $gg \rightarrow b\bar{b}l^+l^-$;
15 diagrams for the reaction $q\bar{q} \rightarrow b\bar{b}\nu\bar{\nu}$;
8 diagrams for the reaction $gg \rightarrow b\bar{b}\nu\bar{\nu}$,

where $l = e, \mu, \tau$ and $q = u, d, s$ quarks.

We will describe a calculation procedure in the next part but here we will formulate the results. The total cross section behavior is presented in Fig. 1 at $\sqrt{s} = 2$ and 4 TeV versus the Higgs boson mass. For comparison the contribution of the Higgs signal diagrams ( processes (1), (2) ) is presented also in the same Fig. 1. One can see that, for instance, at $\sqrt{s} = 2$ TeV the Higgs production cross section is about $0.3 - 0.03$ pb depending on the Higgs mass and therefore one can expect $300 - 30$ signal events for integrated luminosity $L = 1000$ pb$^{-1}$. However the total reaction rates are about two orders of magnitude larger that the Higgs signal. That's why one has to make a detailed analysis of different distributions to find out whether it is possible to apply some set of cuts to suppress the huge background.

Higgs production event rates with the final states (3)-(5) are shown in Table 1 for the energies $\sqrt{s} = 2$ TeV and $\sqrt{s} = 4$ TeV. The number of events corresponds to the integrated luminosity $L = 1000$ pb$^{-1}$. The results for the signal event rates have been obtained by means of PYTHIA 5.6/JETSET 7.3 Monte-Carlo generator [4] and include QCD radiative corrections and the effects of initial and final state radiation.

Various $H^0$ signal event distributions for the both mechanisms (1), (2) are presented in Fig. 2 for the case $M_H = 80$ GeV and $\sqrt{s} = 2$ TeV and in Fig. 3 for $M_H = 120$ GeV and $\sqrt{s} = 4$ TeV. In particular we consider the b-quark and charged lepton transverse momentum distribution — $p_{\perp b}$ and $p_{\perp l}$, the total $b\bar{b}$ pair transverse momentum distribution — $p_{\perp b\bar{b}}$, and missing transverse momentum distribution — $\not{p}_\perp$. The following detailed discussion in section 3 leads us to the conclusion that application of the appropriate cuts on these variables should suppress the overwhelming background to a level small enough.

It is important to point out, that in some cases it is not possible to distinguish signature (7) connected with $WH$ process (1) and signature (8), (9) connected with $ZH$ production mechanism (2). Indeed in the case if a lepton stays in the beam pipe or missed in detector, signature (7) looks like (8) one. If the only one lepton is detected, the signature (9) imitates signature (7). Due to this fact it is reasonable to consider the Higgs signal contribution together. Of course, in this case one needs to simulate the corresponding background as well.

## 3  Background study. Extraction of the Higgs signal.

From the complete set of diagrams mentioned above the dominant contributions to the reactions (3)–(5) come from several parton subprocesses. In Fig. 4. the diagrams for QCD $2 \to 3$ type main background subprocesses

$$\begin{aligned} q\bar{q}' &\to Wb\bar{b} \\ q\bar{q} &\to Zb\bar{b} \\ gg &\to Zb\bar{b} \end{aligned} \quad (11)$$

are presented.

A very important background is connected with $2 \to 2$ electroweak subprocesses

$$\begin{aligned} q\bar{q}' &\to WZ \\ q\bar{q} &\to (Z/\gamma^*)(Z/\gamma^*) \end{aligned} \quad (12)$$

with subsequent leptonic decays of W- and Z-bosons or virtual $\gamma^*$'s.

An important contribution to the background comes from the parton reaction of single top quark production

$$q\bar{q}' \to t\bar{b} \text{ or } \bar{t}b \quad (13)$$

with subsequent decay of top quark $t \to Wb$. In our calculations we have put the top mass equal to 170 GeV. The contribution of the background (13) has not been taken into account in the previous studies [1, 2]. Below we will see that this contribution becomes more important with the growth of the invariant $b\bar{b}$-pair mass.

Contributions of the other diagrams as well as different interference terms from the complete set of Feynman diagrams are small ( about 3–5% ) and one can neglect it in the analysis. The situation here is very similar with 4-fermion final states reactions in $e^+e^-$ collisions $e^+e^- \to b\bar{b}\mu^+\mu^-$, $b\bar{b}e^+e^-$, $b\bar{b}\nu\bar{\nu}$ [5], where also the main contribution comes either from resonant diagrams or from diagrams with soft photons. The reactions (11) were considered in the literature ( $Wbb$ in [6], $Zbb$ in [7]) as well as processes (12) which were calculated with the next to leading order radiative corrections ( $ZZ$ in [8] and $WZ$ in [9]) giving an effect of order 30%.

We present the results of the calculations taking into account decays of W- and Z- bosons with four body final state kinematics. The following procedure has been used in the case of $2 \to 3$ subprocesses (11). Feynman diagrams generation, analytical calculation of the squared matrix elements and generation of the corresponding FORTRAN codes have been done with the help of CompHEP software package [10]. Based on FORTRAN code, a Monte-Carlo generator for event flow at the parton level was made. This generator has been written as a user process subroutine for PYTHIA package which develops the subsequent decay and fragmentation of partons. One of the processes of our generator $gg \to Zb\bar{b}$ was already in PYTHIA, and it has been used as an additional test. For the subprocesses of the $2 \to 2$ type (12) the standard PYTHIA built-in processes have been used for event simulation.

The HMRS set from PDFLIB [11] for the proton structure function have been chosen. However in order to study the level of uncertainties of calculations due to different parameterizations two other structure functions (M-T set B1 and GRV LO) have been used also. As a result we have found that the level of uncertainty is about $25 - 30\%$.

In Fig. 5,6 the results of calculation and simulation of the same distributions as for the signal are presented for background processes:
$p\bar{p} \to WZ \to l\nu b\bar{b}$;
$p\bar{p} \to Wb\bar{b} \to l\nu b\bar{b}$;
$p\bar{p} \to ZZ \to b\bar{b}l\bar{l}(\nu\bar{\nu})$;
$p\bar{p} \to Zb\bar{b} \to b\bar{b}l\bar{l}(\nu\bar{\nu})$
at $\sqrt{s} = 2$ TeV and $\sqrt{s} = 4$ TeV respectively. The main differences in the distributions are connected to the fact that the events of QCD backgrounds (11) (Fig.4) are concentrated most of all in the region of small $p_\perp$. As a result of the detailed comparison of the signal and backgrounds distributions the appropriate set of the optimum cuts has been found:

- $p_{\perp b} > 20$ GeV;   $p_{\perp l} > 15$ GeV ;
- $p_{\perp b\bar{b}} > 20$ GeV ; $\quad\quad\quad\quad\quad\quad\quad\quad\quad\quad\quad\quad\quad\quad\quad\quad\quad\quad\quad\quad\quad\quad$ (14)

the following cuts have been used also for the event selection typical for TEVATRON detectors [2]:

- $\not{p}_\perp > 20$ GeV
- $|\eta_b| < 1.5$;   $|\eta_l| < 2$
- $|\Delta R_{b\bar{b}}| > 0.7$;   $|\Delta R_{bl}| > 0.7$   ($|\Delta R_{b\bar{b},l}| = \sqrt{\Delta \phi^2_{b\bar{b}} + \Delta \theta^2_{b\bar{b},l}}$
- resolution for $M_{b\bar{b}}$:   $\Delta M_{b\bar{b}}/M_{b\bar{b}} = 0.8/\sqrt{M_{b\bar{b}}} + 0.3$, which based on jet energy resolution
$\to 2\Delta M_{b\bar{b}} + M_{b\bar{b}} > M_{b\bar{b}} > -2\Delta M_{b\bar{b}} + M_{b\bar{b}}$ . $\quad\quad\quad\quad\quad\quad$ (15)

This set of cuts reduce the total background more than 40 times. We would like to stress the efficiency of $p_{\perp b\bar{b}}$ cut which has not been used in [2]. This cut improves the ratio of signal to the background by about two times.

The number of events corresponding to signal and background after cuts is presented in the Tables 2,3 for $\sqrt{s} = 2$ TeV and $\sqrt{s} = 4$ TeV respectively. One can see that using 3-standard deviation criteria one can find Higgs up to mass $\simeq 100$ GeV at $\sqrt{s} = 2$ TeV and $\simeq 120$ GeV at $\sqrt{s} = 4$ TeV. These numbers are obtained if we assume that the efficiency of double b-tagging system $\epsilon_b$ is about 50% for $b\bar{b}$ pair registration. A rather high $\epsilon_b$ is very important for the task of the Higgs search, because as it was mentioned in [1, 2] there is a huge background from the events with light jets production: $p\bar{p} \to Wjj + X$;   $p\bar{p} \to Zjj + X$, which has to be suppressed by efficient b-tagging.

One can see that with the decreasing Higgs boson mass the contribution of the $bt$ process (13) to the total background becomes more significant. It should be noted also that $Wbb$ background increases approximately by factor two with the increase of $\sqrt{s}$ from 2 to 4 TeV as well as the $WH$ ( or $ZH$ ) signal. At the same time the $Zbb$ cross section becomes about 3-4 times higher because the $gg \to Zb\bar{b}$ subprocess contribution grows about 6 times. That is why the increasing of the collider energy from 2 to 4 TeV extends Higgs mass interval only up to 120 GeV.

In Fig. 7,8 we demonstrate the effective $b\bar{b}$-invariant mass distribution for the cases $M_H = 80$ GeV, $\sqrt{s} = 2$ TeV and $M_H = 120$ GeV, $\sqrt{s} = 4$ TeV respectively without any cuts and with cuts (14)-(15) applied. All event numbers correspond to $L = 1000$ pb$^{-1}$ and a clear Higgs peak can bee seen in Fig. 7(b) and 8(b).

However the distributions in Fig. 7,8 and event numbers shown in the Table 2,3 have been obtained without real $b-$ quark fragmentation into jets. This question will be considered in the nearest future.

# Conclusions

The present paper continues the study of the Higgs boson search possibility at the TEVATRON energies. In comparison with the results of previous considerations [1,2] the detailed analysis of the 4-fermion final state reactions (3) – (5) is presented. Among the complete set of Feynman diagrams the main contribution to the final states comes from the Higgs signal $WH$ and $ZH$, from the backgrounds $WZ$, $ZZ$ and gluonic background $Wb\bar{b}$, $Zb\bar{b}$ with subsequent decays of $W$, $Z$ bosons. It is shown that the important contribution to the background is connected with single top quark production subreactions $q\bar{q} \to t\bar{b}$ or $\bar{t}b$ with the top decay to $Wb$. Such a background has not been considered in previous papers [1,2]. As a result of various distributions for Higgs signal and background a number of optimal cuts (14), (15) has been found. These cuts suppress background for more than by factor 40 while the signal only about by factor 2.5. The final effective $b\bar{b}$ – pair mass distribution allows to extract the signal from the Higgs with mass up to about 100 GeV at $\sqrt{s} = 2$ TeV and up to about 120 GeV at $\sqrt{s} = 4$ TeV for the integrated luminosity $L = 1000$ pb$^{-1}$.

We should stress also that increase of the collider energy by a factor two extends the Higgs mass interval not more than 20%. This fact is connected with the high magnification of the gluonic background with energy. Therefore for the task of the Higgs search the most important point is the increase of the collider luminosity.

# Acknowledgements

We would like to thank P. Ermolov and P. Grannis for simulating discussions and remarks. We are grateful to M. Dubinin, J. Gunion, E. Shabalina and S. Shichanin for fruitful discussions.

# Figure Captions

Fig.1 – The total cross section for $p\bar{p} \to (l\bar{\nu}, \nu\bar{\nu}, l\bar{l}) + b\bar{b} + X$ processes (including the Higgs boson signal) versus Higgs boson mass for $\sqrt{s} = 2$ and 4 TeV.

Fig.2 – Different $H^0$ signal event distributions for the both mechanisms (1), (2) and $L = 1000$ pb$^{-1}$, for the case $M_H = 80$ GeV and $\sqrt{s} = 2$ TeV.

Fig.3 – Different $H^0$ signal event distributions for the both mechanisms (1), (2) and $L = 1000$ pb$^{-1}$, for the case $M_H = 120$ GeV and $\sqrt{s} = 4$ TeV.

Fig.4 – the diagrams for QCD $2 \to 3$ type main background subprocesses $q\bar{q}' \to Wb\bar{b}$, $q\bar{q} \to Zb\bar{b}$ $gg \to Zb\bar{b}$ which were included into our generator. Note, that diagrams for another quark flavors were also included but are not shown here.

Fig.5,6 — different distributions for background processes (like for the signal) for $\sqrt{s} = 2\ TeV$ and $\sqrt{s} = 4\ TeV$ respectively: $p\bar{p} \to WZ \to l\nu b\bar{b}$, $p\bar{p} \to Wb\bar{b} \to l\nu b\bar{b}$, $p\bar{p} \to ZZ \to b\bar{b}l\bar{l}(\nu\bar{\nu})$, $p\bar{p} \to Zb\bar{b} \to b\bar{b}l\bar{l}(\nu\bar{\nu})$.

Fig.7,8 — the effective $b\bar{b}$-invariant mass distribution for the cases $M_H = 80$ GeV, $\sqrt{s} = 2$ TeV (Fig. 7) and $M_H = 120$ GeV, $\sqrt{s} = 4$ TeV (Fig. 8) respectively without any cuts ( 7(a), 8(a) ) and with cuts (14)-(15) applied ( 7(b), 8(b) ). All events number correspond to the $L = 1000$ pb$^{-1}$.

# Table Captions

Table 1 — Higgs production rates for processes (1) and (2) with signature (3), (4) and (5) at luminosity $L = 1000$ pb$^{-1}$.

Table 2 — Number of 4-fermion final state events for Higgs boson production (process (1), (2)) and background (processes (11)-(13)) for $\sqrt{s} = 2$ TeV with signature (3)-(5) and cuts (14), (15) at $L = 1000$ pb$^{-1}$ and $b\bar{b}$ pair tagging efficiency 50%.

Table 3 — Number of 4-fermion final state events for Higgs boson production (process (1), (2)) and background (processes (11)-(13)) for $\sqrt{s} = 4$ TeV with signature (3)-(5) and cuts (14), (15) at $L = 1000$ pb$^{-1}$ and $b\bar{b}$ pair tagging efficiency 50%.

Table 1:

| $M_H$(GeV) | $pp \to W^{\pm}H \to $ 2b-jet + lepton + $\nu$ | | $pp \to ZH \to $ 2b-jet + 2-lepton | |
|---|---|---|---|---|
| | $\sqrt{s} = 2$ TeV | $\sqrt{s} = 4$ TeV | $\sqrt{s} = 2$ TeV | $\sqrt{s} = 4$ TeV |
| 60  | 345 | 619 | 199 | 355 |
| 80  | 160 | 307 | 89  | 171 |
| 100 | 83  | 170 | 48  | 96  |
| 120 | 42  | 91  | 25  | 55  |
| 140 | 16  | 38  | 10  | 22  |

Table 2:

| $M_H$(GeV) | $WH$ | $ZH$ | $Wb\bar{b} + WZ + b\bar{t}(\bar{b}t)$ | $Zb\bar{b} + ZZ$ |
|---|---|---|---|---|
| 60  | 35 | 27 | 53 + 1 + 8   | 41 + 0 |
| 80  | 27 | 19 | 45 + 43 + 12 | 40 + 28 |
| 100 | 17 | 12 | 35 + 44 + 12 | 32 + 29 |
| 120 | 9  | 7  | 26 + 1 + 13  | 25 + 1 |
| 140 | 4  | 4  | 18 + 0 + 11  | 17 + 0 |

Table 3:

| $M_H$(GeV) | $WH$ | $ZH$ | $Wb\bar{b} + WZ + b\bar{t}(\bar{b}t)$ | $Zb\bar{b} + ZZ$ |
|---|---|---|---|---|
| 60  | 57 | 40 | 82 + 3  + 18 | 122 + 1 |
| 80  | 43 | 30 | 74 + 83 + 15 | 138 + 69 |
| 100 | 29 | 21 | 62 + 83 + 18 | 112 + 71 |
| 120 | 18 | 13 | 46 + 3  + 21 | 88  + 4 |
| 140 | 8  | 7  | 37 + 0  + 20 | 69  + 1 |